\documentstyle[12pt,epsfig]{article}
\begin{document}
\begin{titlepage}
\vskip 0.5cm
\begin{center}
{\Large 
$e^+e^- \to b \bar{b} u \bar{d} \mu^- \bar{\nu}_\mu$ with a $t\bar{t}$ production}
\end{center}
\vskip 0.5cm
\begin{center}
{\Large F. Yuasa\footnote{fukuko.yuasa@kek.jp}, Y.
Kurihara\footnote{kurihara@minami.kek.jp} and S. Kawabata\footnote{kawabata@minami.kek.jp
}} \\
High Energy Accelerator Research Organization\\
Oho 1-1, Tsukuba, Ibaraki 305, Japan
\end{center}

\begin{abstract}
The cross section of $e^+e^- \to b \bar{b} u \bar{d} \mu^-
\bar{\nu}_\mu$ process with a complete set of tree diagrams, 232
diagrams in the unitary gauge, was calculated at the energy range of
$\sqrt{s}$ = 340 - 500 GeV by using {\tt GRACE} system. 
A main contribution to the cross section comes from $t\bar{t}$
production, where $t$ and $\bar{t}$ decay
into $bu\bar{d}$ and $\bar{b} \mu^- \bar{\nu}_{\mu}$, respectively.
It was found that the interference between  the diagrams 
with $t\bar{t}$ production and those with single-$t$ through
$ WW $ pair production amounts to 10\% at the
$t \bar{t}$ threshold energy region.
In the energy region above twice of the top quark mass, more than 95\% of
the cross section comes from the $t\bar{t}$ diagrams. 

\end{abstract}
\end{titlepage}

\section{Introduction}

The top quark physics\cite{top1, top2, top3, top}
is one of the most interesting targets in the near
future $e^+e^-$ linear collider experiments\cite{lc}. 
Since the top quark mass is around 174
GeV\cite{fermi1, fermi2}, 
by the linear colliders at the energy $ {\sqrt{s}} = 340 - 500 $
GeV the top physics  can  be studied precisely.
For example, due to the large top decay width, $t$ decays into $bW $ 
without non perturbative QCD interaction at the long
distance. This means that the study of perturbative QCD can be
done clearly by using the $t\bar{t}$ production process\cite{fadin,peskin}.
\par
So far, the investigation of the top quark physics at the future $e^+e^-$
linear colliders has made a great progress both from the theoretical
and experimental points of view.
At the $t \bar{t}$ threshold region,  the $t\bar{t}$ cross section is
expected to depend on several parameters as follows:
\begin{eqnarray*}
\sigma_{t\bar{t}} ( \sqrt{s}; m_t, \Gamma_t, \alpha_s(m_Z), m_H, \beta_H),
\end{eqnarray*}
where $\beta_H$ is the top YUKAWA coupling. Fujii {\it et al.} have
studied quantitatively on the determination of above physical
parameters under the realistic conditions.
A precise $t\bar{t}$ cross section scan was discussed
including the initial state radiation, beamstrahlung, and beam
energy spread\cite{top1}.
Above the $t\bar{t}$ threshold region, the future $e^+e^-$ linear colliders can provide
opportunities to investigate a new physics related the top quark. 
For example, the studies on the top quark couplings have been made by Comas
{\it et al.}\cite{top2}. 
\par
The $t\bar{t}$ pair production process decays mainly into:
6-jets ($e^+e^- \to b\bar{b}qq'qq'$), 
4-jets plus 1 charged lepton ($e^+e^- \to b\bar{b}qq'l{\nu}$), 
and 2-jets plus 2 charged leptons ($e^+e^- \to b\bar{b}l{\nu}l{\nu}$),
where $q=u,c$, $q'=d,s$, and $l=e, \mu, \tau$.  
In this paper we concentrated on the second,
the semi-leptonic process $e^+e^- \to b \bar{b} u \bar{d} \mu^- \bar{\nu}_\mu$.
This semi-leptonic (4-jets plus 1 charged lepton) process has several
advantages in
the event selection for the
top quark physics both at the $t \bar{t}$ threshold region and above.
Firstly, since the semi-leptonic process
has less than 6-jets, it is less suffered from the
backgrounds due to the wrong combination of jets.  
Secondly, these semi-leptonic events are produced with enough
statistics because 
$Br(t\bar{t} \to b\bar{b}qq'l\nu)$ is about 44\% and 
$Br(t\bar{t} \to b\bar{b}qq'qq')$ is 45\% after QCD correction. 
Moreover, when we use the electric charge of the lepton as tag,  
we can identify the
hemisphere where 3-jets, the decay products of the top quark,  
appear.
\par 
Although the main contribution to the total
cross section at $\sqrt{s}$ = 340  - 500 GeV comes from the
diagrams through the $t \bar{t}$ production,
those from the other diagrams are not negligible small at all.
To study the $t\bar{t}$ production process precisely, an accurate
estimation of the other diagrams
is essentially required. 
We have calculated the cross
section of  $e^+e^- \to b \bar{b} u \bar{d} \mu^- \bar{\nu}_\mu$
process with a full set of Feynman diagrams, 232 diagrams
in unitary gauge. 
Generally speaking, as the number of final state particles increases, 
the calculation of the cross section becomes more complicated and more
tedious because there exist a lot of Feynman diagrams
and there occur interferences among these diagrams.
In order to cope with this complex calculation, we have used an
automatic Feynman diagrams calculation package, {\tt
GRACE}\cite{grace}.
All the results obtained for the process $e^+e^- \to
b\bar{b}u\bar{d}\mu^- \bar{\nu}_{\mu}$ are immediately applicable to the
other semi-leptonic processes such as $e^+e^- \to b\bar{b}c\bar{s}
\mu^- \bar{\nu}_\mu$ when we neglect the difference of mass between
$u,d$ and $c,s$. 
\par
In addition  to an accurate calculation of the total cross section,
the effects of interferences among diagrams have been studied.
These interferences change 
the $t {\bar{t}}$ threshold shape destructively.
Also we present the invariant mass and angular distributions of
the final state particles at $\sqrt{s}$=500 GeV. 
In order to enhance the signals of $t\bar{t}$ production process
against the background, 
we studied on the effective  
cuts in the invariant mass distributions such as $M_{bu\bar{d}}$ and $M_{b\bar{b}}$
and the cut for the angle of the final lepton in these distributions.
Recently, the similar calculations of $e^+e^- \to$ six fermions are
performed by  two groups. Montagna {\it et al.}\cite{mont}
reported a full calculation of the six-fermion process
in the $e^+e^-$ linear colliders, 
where the higgs boson  with the intermediate mass  are produced, 
by the program package {\tt ALPHA}\cite{alpha} for the matrix elements
calculations and 
{\tt HIGGSPV/WWGENPV}\cite{higgspv} for Monte Carlo event generation.
Accomando {\it et al.}\cite{ball} have been calculating a semi-leptonic process such as
$e^+e^- \to b\bar{b}qq'l\nu$ by program package {\tt PHACT}\cite{phact}.
\par
In Section 2, we present the details of the computational parameters
for the calculations in {\tt GRACE} at first.
Secondly, the numerical results are shown: the total cross sections
(Section 2.1), effects of the interferences among diagrams (Section
2.2), and the mass and angular distributions of interest (Section 2.3).
\par
\section{Cross Section Calculation}
The cross sections have been calculated based on a complete set of the
tree level diagrams using {\tt GRACE} system, 
a program package for an automatic Feynman diagram calculation and
the event generation.
When the initial and final states are specified, {\tt GRACE} generates the
matrix elements in terms of helicity amplitudes. 
The total cross section is computed by the Monte
Carlo integration of the matrix elements over the phase space\cite{bases}.
As for the kinematics, we only have to
prepare the suitable program code to the physics process of interest.  
Usually, as the number of the final state particles increases, 
it becomes more and more difficult to prepare such a program code of
kinematics that may catch every singularities appeared in all diagrams
and give a good convergence of the multidimensional Monte Carlo integration.
We have developed a kinematics code for the process  
$e^+e^- \to b \bar{b} u \bar{d} \mu^- \bar{\nu}_\mu$, which makes 
the calculation of the total cross sections converge rapidly enough to a
good accuracy of better than 1 \% at ${\sqrt{s}} = 340 - 500 $ GeV. 
\par
The cross sections has been calculated with the parameters shown in Table 1.
The widths of $W$ and $Z$ bosons are taken to be the fixed values.
The QCD correction is not included in the calculations.
The quark masses are set as in Table 2.
The $t$ quark width of 1.558 GeV corresponds to
the decay width of $t \to b W$, which means
that the decay branching ratio of $t \to b W$ channel is 100\%.
\par
The total cross section of  $e^+e^- \to t \bar{t}$ process is calculated
to be $5.77 \times 10^{-1}$ pb  at CM energy of 500 GeV by {\tt GRACE}. 
This is consistent with
the cross section of  $e^+e^- \to b \bar{b} W^+ W^-$ processes of
$5.78 \times 10^{-1}$ pb calculated by {\tt GRACE}. The cross section
of 
$e^+e^- \to b \bar{b} u \bar{d} \mu^- \bar{\nu}_\mu$ 
process only with the $t \bar{t}$ production diagrams is calculated to be
$2.13 \times 10^{-2}$ pb ( in Table 3) and is also consistent with 
$\sigma_{t \bar{t}} \times Br(W \to u \bar{d}) \times Br(W \to \mu^-
\bar\nu^{\mu} ) =2.14 \times 10^{-2}$ pb. 
Here, we took $Br(t \to bW)$ = 100 \%.
\par
\subsection{Results of the numerical calculation}

Among all 232 diagrams (unitary gauge) of the process, 
the $t \bar{t}$ production diagrams of Fig.1 are dominant
over the others. 
We have divided the {\it major background diagrams} into three categories:
the diagrams with $W^+ W^- \gamma$ (hereafter $WW\gamma$) in Fig.2,
those with $W^+ W^- Z$ (hereafter $WWZ$)
in Fig.3, and those with single-$t$ through $W^+W^-$ pair production 
 (hereafter $tWW$) in Fig.4.  
\par
The total cross sections with a full set of  diagrams 
at the CM energy points of 340 - 500 GeV are summarized in Table 3 and
Fig.5 {\tt a)} and {\tt b)}.
Both in {\tt a)} and {\tt b)} of Fig.5, solid lines show the
numerical result of the total cross section with all the diagrams. 
The result with only the dominant 
$t \bar{t}$ diagrams are shown by dashed line in both {\tt a)
} and {\tt b)}.
Besides results with all the diagrams and that with
$t \bar{t}$ diagrams, we also showed those with $t \bar{t}$ and the major
background diagrams. 
Dotted lines in {\tt a)} and {\tt b)} show the result with $t
\bar{t}$  and $ WW \gamma$ and that with $t \bar{t}$ and $ WWZ$,  
respectively.
Dot-dashed lines in {\tt a)} and {\tt b)} show the result with 
$t \bar{t}$, $ WW \gamma$, and $tWW$
and that with $t \bar{t}$, $ WWZ $, and $tWW$, respectively.
These results shown by dotted lines and dot-dashed lines include the
interferences among selected diagrams.
\par
As shown in Fig.5, cross sections with both $t \bar{t}$ and the major
background diagrams show different behaviors from that with all 
the diagrams.
The difference between them is about 3\% at $\sqrt s$ = 500 GeV.
This means that the effect from the interference between $t \bar{t}$
and the rest diagrams except $WW\gamma$, $WWZ$, and $tWW$ are also
not negligible and important.
The contribution from the background diagrams to the total
cross section, ${(\sigma_{all} - \sigma_{t \bar{t}})}/{\sigma_{t\bar{t}}}$, 
is less than 5\% in total above 
the energy of twice of the top quark mass. 
\par
\subsection{Interference between diagrams}
Due to the large decay width of the top quark, at the  $t \bar{t}$ threshold
region it decays so immediately that the non perturbative
QCD effect at the long distance does not take part.
Thus, the cross section of $ \sigma_{t \bar{t}}$ can be
calculated by means of the perturbative QCD theory\cite{fadin}.
By measuring the cross section of the $t \bar{t}$ production around the 
$t \bar{t}$ threshold region at the future linear colliders, 
a clear test of the perturbative QCD theory is possible.  
To determine the QCD parameters like $\alpha_{s}$, 
it is essential to measure of the cross section accurately.
In our calculation, the effects from initial state
radiation, beamstrahlung, and beam energy spread were neglected.
These effects to the shape of $t\bar{t}$ cross section has been well
investigated by Fujii {\it el al.}\cite{top1, top2, top3}.
Besides these effects, it is important to take 
the effect such as the interferences between $t \bar{t}$ diagrams
and the other background diagrams into account.
\par
It is found that there are
destructive interferences between diagrams with  $t {\bar{t}}$ production
and those with single-$t$ through WW pair production around the $t \bar{t}$
threshold region.
In order to see the behavior of these destructive interferences the values of 
${\sigma_{t\bar{t} + tWW}}/{(\sigma_{t\bar{t}} +
\sigma_{tWW})}$ are plotted at several energy points in Fig.6.
The magnitude of the destructive interference is as large as 10 \% at 
$\sqrt {s}$ = 340 GeV.
Then it drops rapidly as CM energy goes up.
In the energy region above twice of the top quark mass, this destructive
interference becomes less than 2 \% and above $\sqrt{s}$ = 360 GeV it becomes 
negligible small.

\par
When we generate Monte Carlo events with the luminosity 1 ${\rm
fb}^{-1}$ in each energy points, the destructive interference of 10 \%
is not negligible compared to the statistical errors at the $t\bar{t}$
threshold region.
For example, at $\sqrt{s}$ = 350 GeV
the total cross section of $\sigma_{t \bar{t}}$ is $ 1.90
\times 10^{-1}$ pb and the statistical error is estimated to be a few \%.
\par

\subsection{Invariant mass and angular distributions}

Above the $t \bar{t}$ threshold energy region, there open the opportunities of
measuring the top production and decay couplings. 
For this analysis,  the process of semi-leptonic 4-jets plus 1 charged
lepton is useful to distinguish $t$ from $\bar{t}$ by the electric charge of the lepton. 
Generally, in order to select the semi-leptonic $t\bar{t}$ production
events,  firstly
an isolated charged lepton and 4 jets are required in the event. 
Then two jets out of the four are to be identified as $b$ jets in the
vertex detector. 
From the electric charge of lepton, we can determine the hemisphere
where 3 jets from the top quark appear. 
If the invariant
mass of 2 jets in this hemisphere and one of $b$ jets becomes nearly equal to the mass of the top quark, we
accept it as the $t\bar{t}$ production event. 
\par
The invariant mass distribution of $b$, $u$ and
$\bar{d}$, and that of $b$ and $\bar{b}$ 
at $ \sqrt{s} = 500$ GeV are shown in Fig.7 {\tt a)} and {\tt b)}, respectively.
In Fig.7, histograms show the invariant mass distributions for the diagrams
of $t \bar{t}$ production and crosses show
those for all the diagrams.
As shown in Fig.7 {\tt a)}, if we take only those
events around the
top quark mass on the  $M_{bu\bar{d}}$
distribution, the contribution from the background diagrams becomes
negligibly small.
At the energy of $\sqrt{s}$ = 500GeV, when the top quark mass cut,
{\it e.g.}
148 GeV $\leq M_{bu\bar{d}} \leq$ 204 GeV is applied, the
contribution from the background diagrams is reduced from about 5\% to less than 3\%.
In Fig.7 {\tt b)}, crosses shows that the contribution from
$WW\gamma$ diagrams can be seen at the low-end 
and the contribution from $WWZ$ diagrams can be seen at the mass
of Z boson.
\par
For the analysis of the top couplings, it should be noticed that
the top quark decays immediately before it hadronizes and its spin
information 
is transferred to the final state particles\cite{top2}. 
The angles of the final state particles are important parameters in order to
reconstruct the production angle of the top quark.
Here, we show the angular distributions of final $\mu^{-}$ in Fig.8.
The backgrounds in the forward angle region, 
from all the diagrams are significant in the distribution of the 
$\mu^-$ angular distribution, at $ \sqrt{s} = 500$ GeV.
For example, when  the top quark mass cut such as
148 GeV $\leq M_{bu\bar{d}} \leq$ 204 GeV is applied,
the effects from the background
in the angular distribution of $\mu^-$ can be reduced to less than 2\%
but still exists as shown in Fig.8 {\tt b)}.

\section{Conclusion}

The total cross section of 
$e^+e^- \to b \bar{b} u \bar{d} \mu^- \bar{\nu}_\mu$ process 
at the energy range of $\sqrt{s}$ = 340 - 500 GeV 
was calculated with a complete set of tree diagrams by using
{\tt GRACE}. 
Among all diagrams, the main contribution to the total cross section
comes from the  $t\bar{t}$ production 
diagrams. 
The background to the 
$t{\bar{t}}$ production from all the diagrams is below 5\% above 
the $t{\bar{t}}$ threshold. 
The contributions from the major background diagrams such as
$WW\gamma$, $WWZ$, and single-$t$ through $WW$ pair production are also calculated.
Compared these results to those with all diagrams, it is found that
the interference between $t \bar{t}$ and  the rest diagrams except the major backgrounds is not negligible.
It is also found that the diagrams of  $t {\bar{t}}$ production and
those with single-$t$ through WW pair
production make a strong destructive interference (as large as 10\%)
around the $t {\bar{t}}$ threshold region. 
This destructive interference has a deep effect on the shape
of the $t {\bar{t}}$ cross section at the $t\bar{t}$ threshold region. 
\par
We calculated the distributions of the invariant masses and the angular
distribution of the final state particles at $\sqrt{s} = 500$ GeV. 
Angular distributions of $\mu^{-}$ shows a large effect of 
backgrounds at the forward angle. 
Even when we apply the cut in the invariant mass distribution of $b$,
$u$, and $\bar{d}$ to enhance the top quark signal, 
the background from the diagrams with $WW\gamma$, 
$WWZ$ and single-$t$ through WW pair did not disappear in 
the angular distribution of $\mu^-$ 
\par
As a conclusion, for the analysis of the top quark physics around the
$t \bar{t}$ threshold region and above, the accurate calculation including full
set of 232 Feynman diagrams is essential.

\par
\section*{Acknowledgements}
The authors wish to acknowledge to Minami-Tateya Collaboration.
Especially, we are indebted to Prof. Y.Shimizu, Prof. T.Ishikawa and 
Dr. J.Fujimoto for valuable suggestions and encouragements.
The authors also wish to thank Dr. E. Accomando and Prof. A.
Ballestrero for useful discussions and comparisons with their work are
gratefully acknowledged.
One of the authors (F.Y.) thanks Dr. T.Omori for fruitful discussions. 
This work is supported in part by Ministry of Education, Science, and
Culture, Japan under Grant-in-Aid for International Scientific Research
Program (No.09044359). 

\newpage
\begin{center}
\normalsize
\begin{tabular}{|c|c|} \hline
$m_Z$ & 91.187 GeV \\ \hline
$\Gamma_Z$ & 2.49 GeV  \\ \hline
$m_W$ & 80.22 GeV \\ \hline
$\Gamma_W$ & 2.052 GeV \\ \hline
${\rm sin}^2\theta_W$ & $1-\frac{m^2_W}{m^2_Z}$ \\ \hline
$\alpha$ & 1/128.07 \\ \hline
\end{tabular}
~\\
~\\  
\normalsize
Table 1  The parameters
\end{center}
\begin{center}
\normalsize
\begin{tabular}{|c|c|} \hline
$m_t$ & 174 GeV \\ \hline
$\Gamma_t$ & 1.558 GeV \\ \hline
$m_b$ & 4.1 GeV \\ \hline
$m_u$ & 2 MeV \\ \hline
$m_d$ & 5 MeV \\ \hline
\end{tabular}
~\\
~\\  
\normalsize
Table 2  The mass and width parameters
\end{center}

\begin{center}
\tiny
\begin{tabular}{|c|c|c|c|c|c|c|} \hline
${\sqrt s}$ & all diagrams & $t\bar{t}$ & $t\bar{t}+WW\gamma$ 
&$t\bar{t}+WW\gamma$+$tWW$
& $t\bar{t}+WWZ$ &$t\bar{t}+WWZ$+$tWW$ \\ \hline
340&$6.87(2)\times10^{-4}$&$4.462(3)\times10^{-4}$&
    $4.936(9)\times10^{-4}$&$6.80(2)\times10^{-4}$& 
    $5.91(1)\times10^{-4}$&$7.80(2)\times10^{-4}$     \\   \hline
350&$6.45(1)\times10^{-3}$&$6.187(4)\times10^{-3}$&
    $6.235(6)\times10^{-3}$&$6.427(9)\times10^{-3}$&
    $6.354(9)\times10^{-3}$&$6.57(1)\times10^{-3}$     \\   \hline
360&$1.497(2)\times10^{-2}$&$1.463(1)\times10^{-2}$&
    $1.467(1)\times10^{-2}$&$1.494(2)\times10^{-2}$&
    $1.483(2)\times10^{-2}$&$1.509(2)\times10^{-2}$     \\   \hline
370&$1.906(4)\times10^{-2}$&$1.864(1)\times10^{-2}$&
    $1.869(2)\times10^{-2}$&$1.902(3)\times10^{-2}$&
    $1.887(3)\times10^{-2}$&$1.919(3)\times10^{-2}$     \\   \hline
380&$2.142(4)\times10^{-2}$&$2.100(1)\times10^{-2}$&
    $2.104(2)\times10^{-2}$&$2.145(3)\times10^{-2}$&
    $2.127(4)\times10^{-2}$&$2.164(3)\times10^{-2}$     \\   \hline
390&$2.298(5)\times10^{-2}$&$2.244(1)\times10^{-2}$&
    $2.247(2)\times10^{-2}$&$2.290(4)\times10^{-2}$&
    $2.268(4)\times10^{-2}$&$2.316(3)\times10^{-2}$     \\   \hline
400&$2.383(5)\times10^{-2}$&$2.329(1)\times10^{-2}$&
    $2.332(2)\times10^{-2}$&$2.385(4)\times10^{-2}$&
    $2.364(5)\times10^{-2}$&$2.409(4)\times10^{-2}$     \\   \hline
450&$2.424(4)\times10^{-2}$&$2.344(2)\times10^{-2}$&
    $2.354(3)\times10^{-2}$&$2.432(4)\times10^{-2}$&
    $2.397(4)\times10^{-2}$&$2.467(5)\times10^{-2}$     \\   \hline
500&$2.232(4)\times10^{-2}$&$2.130(1)\times10^{-2}$&
    $2.150(3)\times10^{-2}$&$2.242(5)\times10^{-2}$&
    $2.216(5)\times10^{-2}$&$2.308(6)\times10^{-2}$     \\   \hline
\end{tabular}
~\\
~\\  
\normalsize
Table 3  The total cross sections in pb.
\end{center}
\newpage
\begin{figure}[h]
\begin{center}
\mbox{\epsfig{file=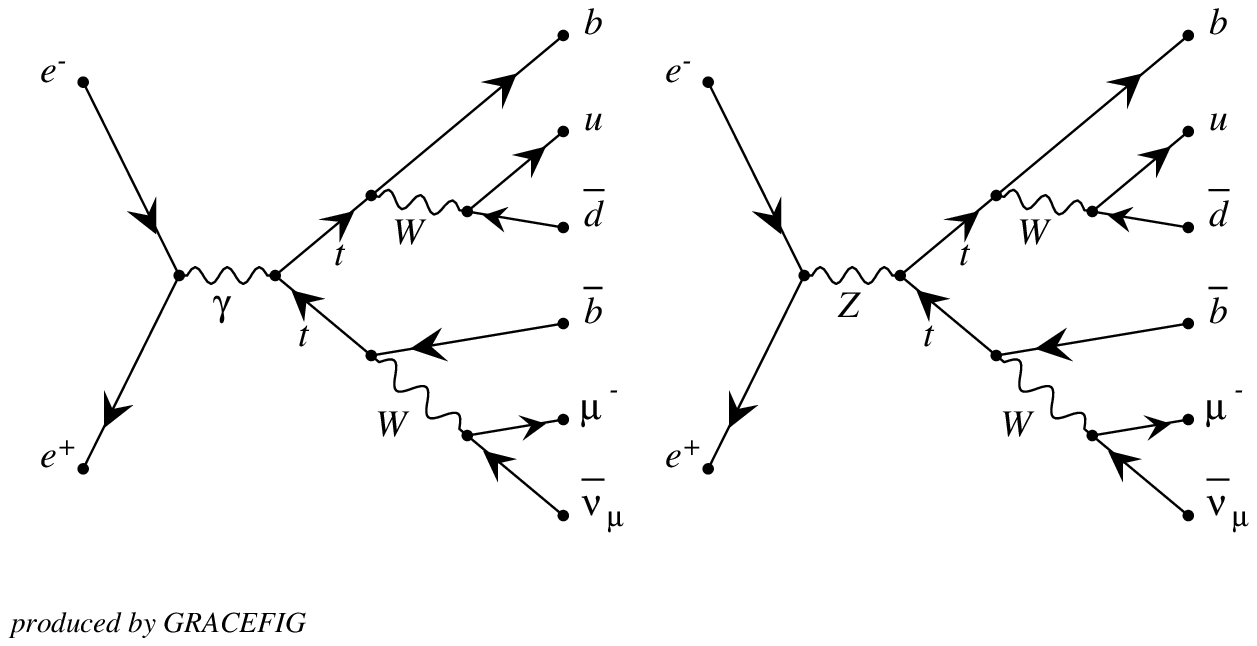,height=10cm}}
\caption{A part of Feynman diagrams for the 
$e^+e^- \to b \bar{b} u \bar{d} \mu^- \bar{\nu}_\mu$ 
process. Diagrams with the $t \bar{t}$ production are shown.
}
\end{center}
\end{figure}
\begin{figure}[h]
\begin{center}
\mbox{\epsfig{file=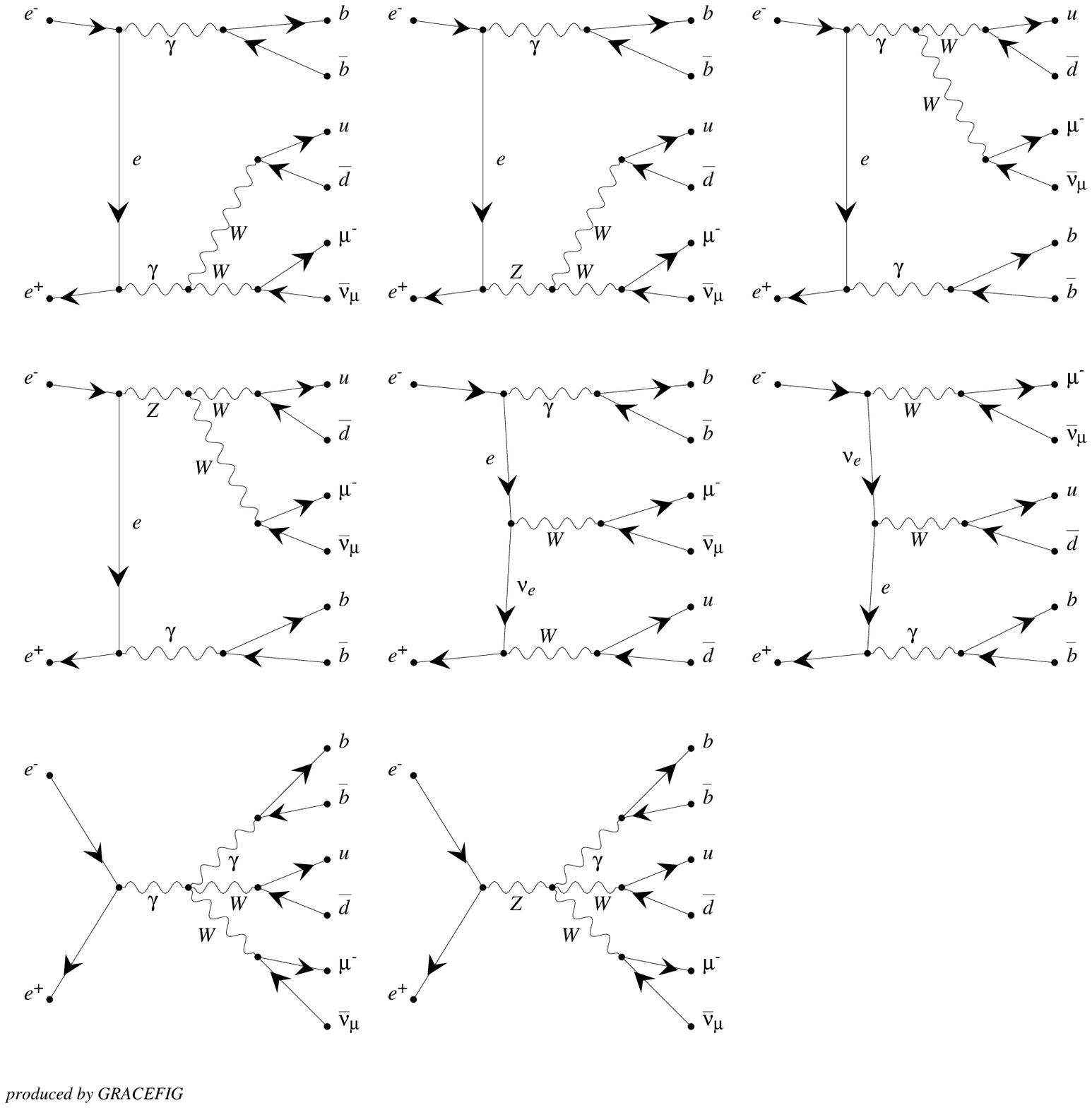,height=14cm}}
\caption{A part of Feynman diagrams for the 
$e^+e^- \to b \bar{b} u \bar{d} \mu^- \bar{\nu}_\mu$ 
process. Diagrams with the $W^+ W^- \gamma$ production are shown.
}
\end{center}
\end{figure}
\begin{figure}[h]
\begin{center}
\mbox{\epsfig{file=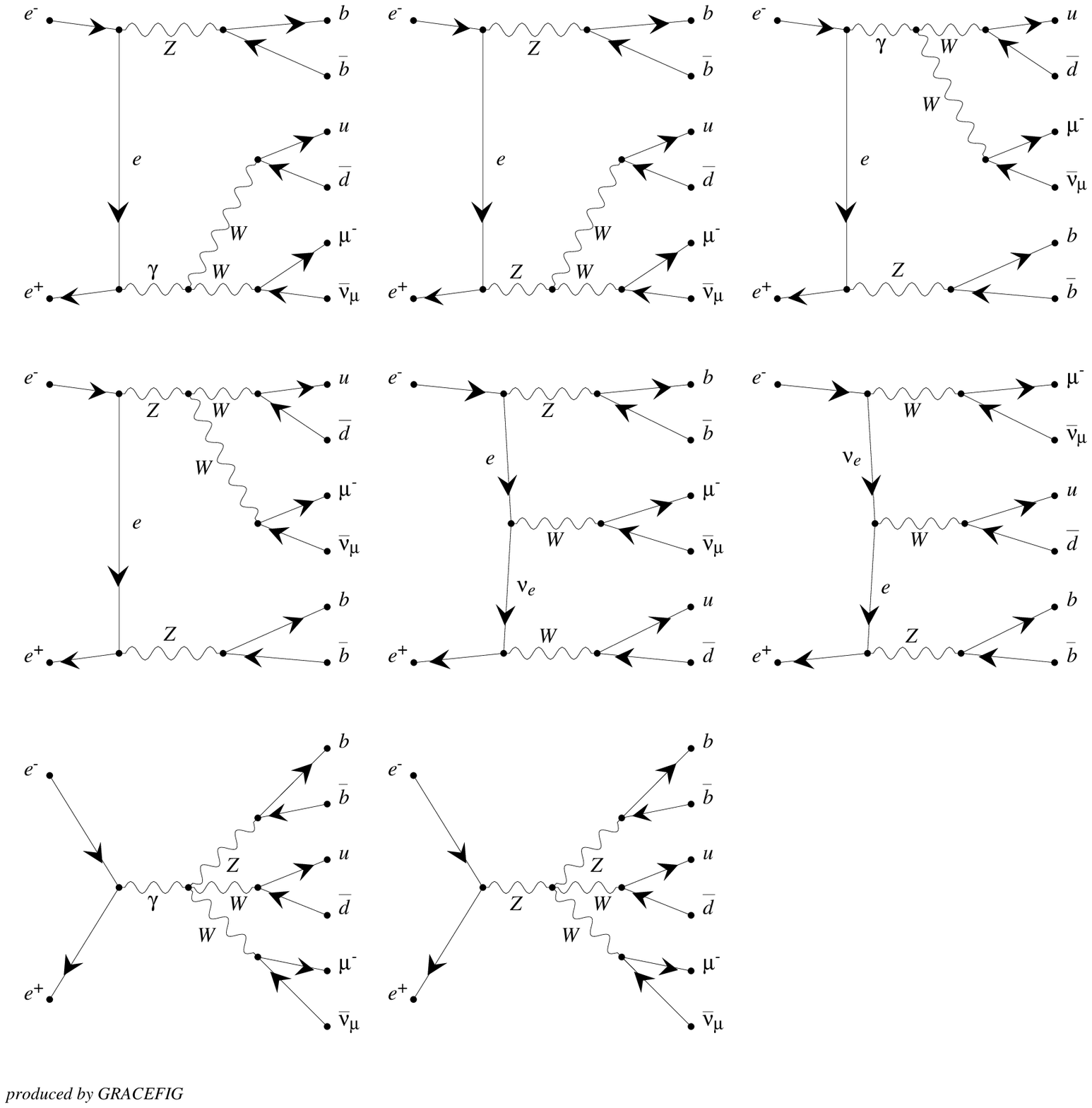,height=14cm}}
\caption{A part of Feynman diagrams for the 
$e^+e^- \to b \bar{b} u \bar{d} \mu^- \bar{\nu}_\mu$ 
process. Diagrams with the $W^+ W^- Z$ production are shown.
}
\end{center}
\end{figure}
\begin{figure}[h]
\begin{center}
\mbox{\epsfig{file=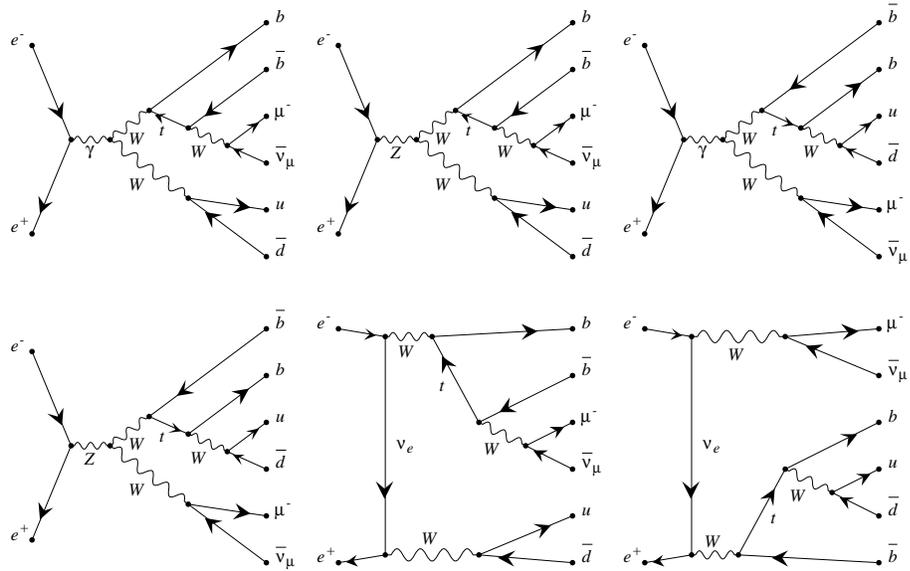,height=14cm}}
\caption{A part of Feynman diagrams for the 
$e^+e^- \to b \bar{b} u \bar{d} \mu^- \bar{\nu}_\mu$ 
process. Diagrams with the single $t$ production through $WW$ pair production
are shown.
}
\end{center}
\end{figure}
\begin{figure}
\begin{center}
\mbox{\epsfig{file=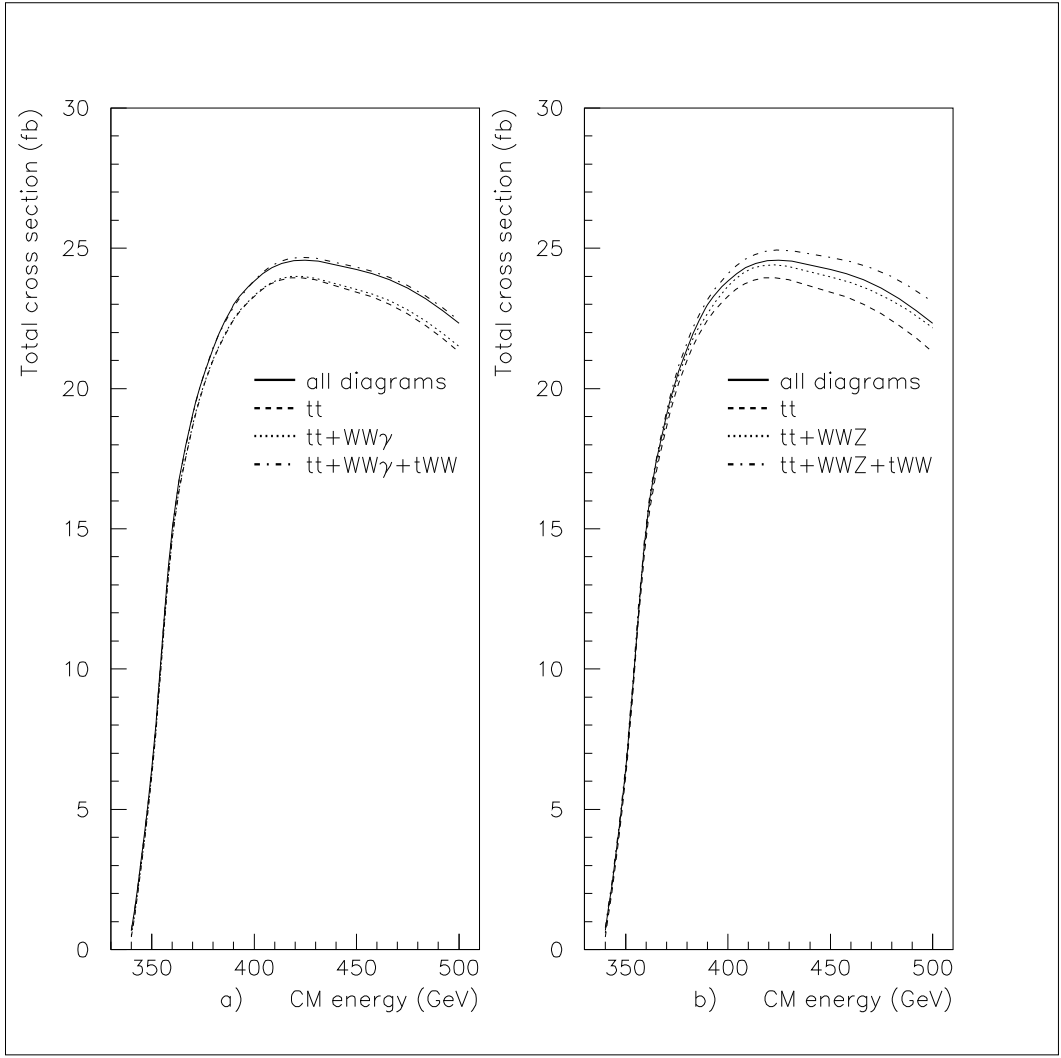,height=14cm}}
\caption{
{\tt a)} The total cross sections with all diagrams (solid line),
with $t \bar{t}$ diagrams (dashed line), with $t \bar{t}$ and 
$WW\gamma$ (dotted line), and with $t \bar{t}$, 
$WW\gamma$, and $tWW$ (dot-dashed line).
{\tt b)} The total cross sections with all diagrams (solid line),
with $t \bar{t}$ diagrams (dashed line), with $t \bar{t}$ and 
$WWZ$ (dotted line), and with $t \bar{t}$,
$WWZ$, and $tWW$ (dot-dashed line).
}
\end{center}
\end{figure}
\begin{figure}
\begin{center}
\mbox{\epsfig{file=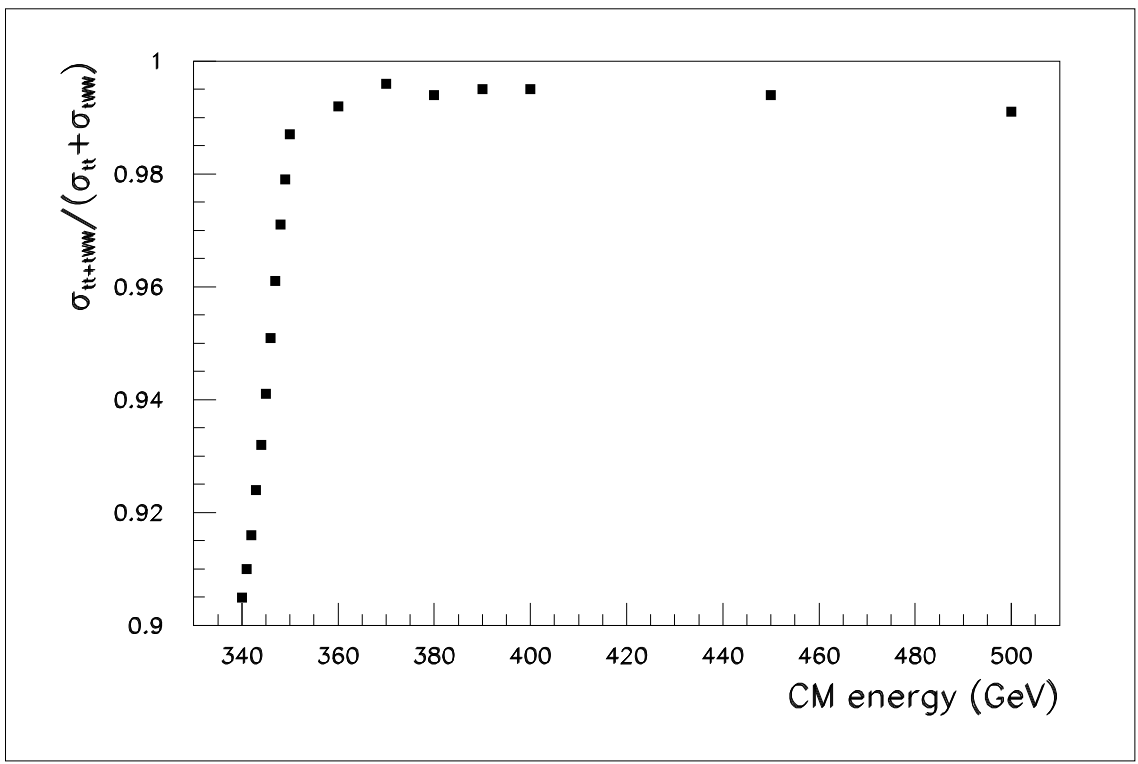,height=10cm}}
\caption{The effect of interference between diagrams with 
$t \bar{t}$ production and ones with single-$t$ through WW pair production.
}
\end{center}
\end{figure}
\begin{figure}
\begin{center}
\mbox{\epsfig{file=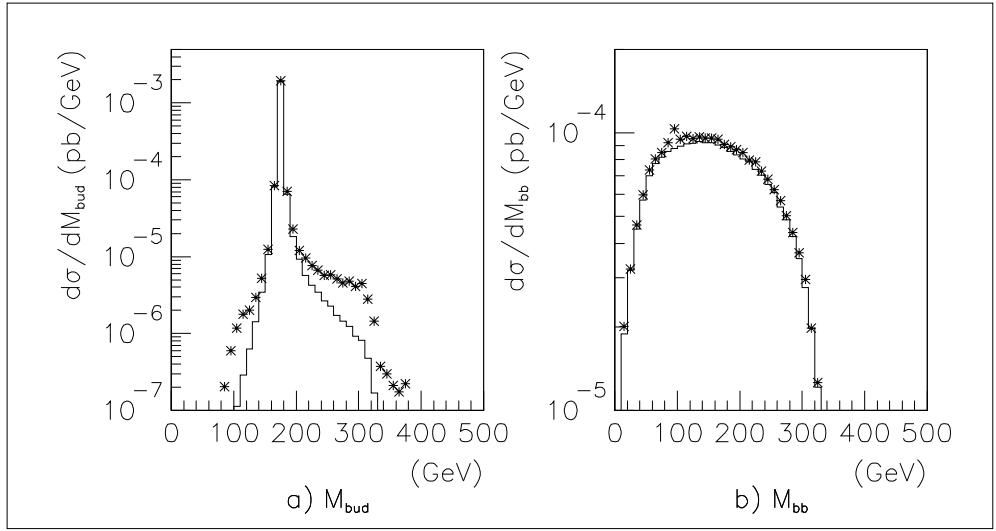,height=7cm}}
\caption{Invariant mass distributions of {\tt a)} $bu\bar{d}$ and {\tt
b)} $b\bar{b}$. Histograms show the contributions from $t\bar{t}$ diagrams only.
Crosses show the contributions from all the diagrams.
}
\end{center}
\end{figure}
\begin{figure}
\begin{center}
\mbox{\epsfig{file=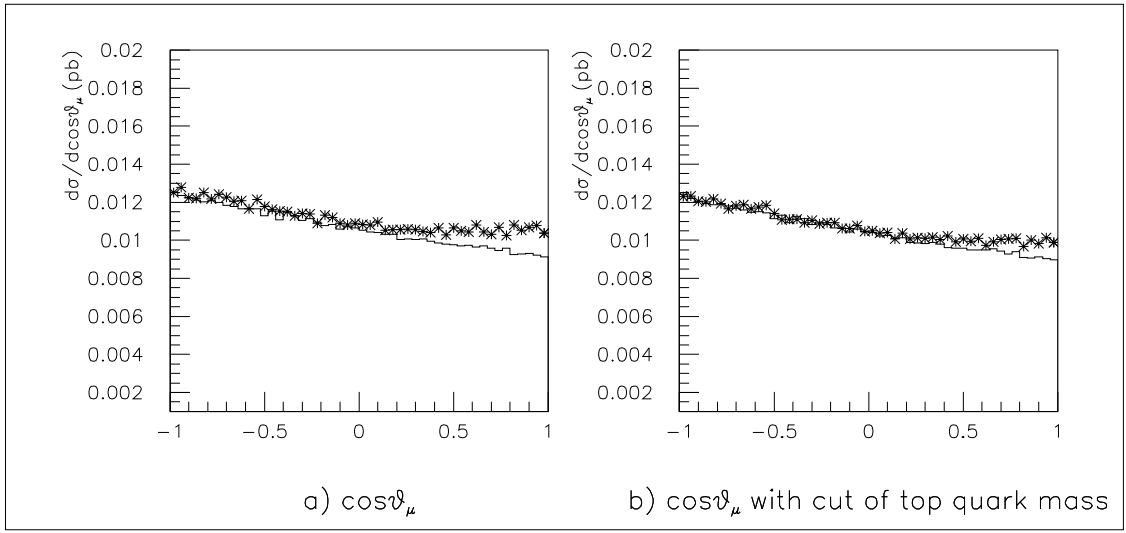,height=7cm}}
\caption{Both {\tt a)} and {\tt b)} show the angular distributions of $\mu$. 
{\tt b)} shows the angular distribution of $\mu$ with the cut of the top quark mass.
Histograms show the contributions from $t\bar{t}$ diagrams only.
Crosses show the contributions from all the diagrams.
}
\end{center}
\end{figure}
\end{document}